\documentclass[preprint]{sigplanconf}

\usepackage{amsmath}

\usepackage{graphicx}

\begin{document}

\setlength{\pdfpageheight}{\paperheight}
\setlength{\pdfpagewidth}{\paperwidth}

\conferenceinfo{CONF '15}{October 27, 2015, Pittsburgh, Pennsylvania, USA} 
\copyrightyear{2015} 
\copyrightdata{978-1-nnnn-nnnn-n/yy/mm} 
\doi{nnnnnnn.nnnnnnn}



\permissiontopublish             

\preprintfooter{Dino Implementation}   

\title{Implementation of the Programming Language Dino -- A Case Study in Dynamic Language Performance}

\authorinfo{Vladimir N. Makarov}
           {Red Hat}
           {vmakarov@gcc.gnu.org}

\maketitle

\begin{abstract}


  The article gives a brief overview of the current state of programming
language Dino in order to see where its stands between other dynamic
programming languages.  Then it describes the current implementation,
used tools and major implementation decisions including how to
implement a stable, portable and simple JIT compiler.

We study the effect of major implementation decisions on the
performance of Dino on x86-64, AARCH64, and Powerpc64. In brief, the performance
of some model benchmark on x86-64 was improved by \textbf{3.1}
times after moving from a stack based virtual machine to
a register-transfer architecture, a further \textbf{1.5} times by adding byte code
combining, a further \textbf{2.3} times through the use of JIT, and a further \textbf
{4.4} times by performing type inference with byte code specialization, with a resulting overall performance improvement of about \textbf{47} times. To put
these results in context, we include performance comparisons
of Dino with widely used implementations of Ruby, Python 3, PyPy
and JavaScript on the three platforms mentioned above.

  The goal of this article is to share the experience of Dino
implementation with other dynamic language implementors in hope that
it can help them to improve implementation of popular dynamic
languages to make them probably faster and more portable, using less
developer resources, and may be to avoid some mistakes and wrong directions
which were experienced during Dino development.

\end{abstract}

\category{D.3.4}{Processors}{Interpreters}

\terms
Languages, Performance, Algorithms

\keywords
Dynamic Language, Language Design, Interpreters, Optimizations, JIT

\section{Introduction}

  Programming language Dino has a long history.  Originally it was
designed, implemented, and used as a simple dynamically typed
scripting language in a small computer game company.  During its long
life, the language and its implementation were dramatically changed.
The implementation was changed many times by using new tools, new
virtual machines, by adding new optimizations.  Therefore its
development could be a good case study of dynamic language
implementation, simple, portable, and with a performance competitive
with JIT implementations of other popular dynamic languages.



  The first part of the article contains brief overview of the current
version of the programming language Dino.
We describe the basic design of the language, its type system and particular features
such as multithreading, heterogeneous extensible arrays, array slices, associative tables,
first-class functions, pattern-matching, as well as Dino's unique approach to class
inheritance via the `use' class composition operator.

  The second part of the article describes Dino's implementation.
We outline the overall structure of the Dino interpreter and just-in-time compiler (JIT) and the design of the byte code
and major optimizations. We also describe implementation details such as the garbage collection system, the algorithms underlying Dino's data structures,
Dino's built-in profiling system,
and the various tools and libraries used in the implementation. Our goal is to give an overview of the major implementation decisions
involved in a dynamic language, including how to implement a stable and portable JIT.

  The third part studies the effect of design choices and major
optimizations on Dino's performance on x86-64, AARCH64, and Powerpc64.
We examine the choice to use a register-transfer virtual machine instead of a
stack-based design and the effects of byte code combining, just-in-time compilation,
and type inference with byte code specialization.

  The fourth part compares Dino's performance with other
popular dynamic language implementations, in order to better understand
the combined effect of all described optimizations. Performance results
are given for the same three platforms as in the previous part.

  We conclude by giving possible directions of further
research based on Dino, whether by testing new optimizations
or studying ways to improve the JIT.

\section{Language Overview}

In this section we describe 
 the current version of the Dino
programming language.
This section does not aim to give a formal or comprehensive specification, and there is a possibility that the omitted details might raise questions for the reader. However, our goal is to give a high-level language overview in order to see how the features of the language compare to those of similar
dynamic languages and to understand how 
the implementation methods used for Dino are applicable to other languages. 


The best way to proceed will be to give a series of small example programs.  To start with, the following is a
Dino implementation of the Sieve of Eratosthenes:

{\footnotesize
\begin{verbatim}
 1.  val SieveSize = 8191;
 2.  var i, prime, k, count = 0, flags = [SieveSize : 1];
 3.  for (i = 0; i < SieveSize; i++)
 4.    if (flags[i]) {
 5.      prime = i + i + 3;
 6.      k = i + prime;
 7.      for (;;) {
 8.        if (k >= SieveSize)
 9.          break;
10.        flags[k] = 0;
11.        k += prime;
12.      }
13.      count++;
14.    }
15.  putln (count);
\end{verbatim}
}

An initial design goal was to make Dino resemble C whenever reasonable
in order to reduce the learning curve for programmers familiar with C.
According to a commonly used classification of programming languages,
Dino can therefore be thought of as a \emph{curly-bracket language}.

Variables in Dino can hold values of any type, so Dino
is a \emph{dynamically typed language}.  The scalar value types
include \emph{char} (Unicode character),
\emph{integer} (64-bit signed integers),
\emph{long} (multi-precision integers) and
\emph{float}\footnote{Dino floats are IEEE 754-1985 double precison
  floating point numbers.}.
Dino types are themselves values\footnote{The type of type values is denoted by the keyword \emph{type}.}.

Variables in Dino should be declared (lines 1 and 2). 
The declaration
scope of a variable starts from the declaration point and finishes at the end of
the block containing the declaration\footnote{A block is a series of statements and declarations enclosed in curly-brackets.}.
It can finish earlier if the identifier is redeclared in the block.
The declaration scope includes nested blocks but excludes scopes of other
declarations with the same identifier in the nested
blocks\footnote{Scope rules in Dino were originally more relaxed,
permitting an identifier to be referenced before its declaration,
 but
they were changed to the current ones when an interactive REPL environment
was implemented.}.

If a variable is declared by a \emph{val} declaration (line 1), then it is a constant
whose value cannot be changed.  

\subsection{Array, Slices, and Tables}

The structured value types include \emph{arrays}.  In the above
example, the variable {\tt flags} holds an array of integers of size {\tt
SieveSize}.  Each element of the array is initialized to integer value 1
(line 2).  Different elements of an array can hold values of different
types.  Arrays are \emph{mutable} by default, i.e. their elements can be
modified, removed, and new elements can be inserted.  An array can be
transformed to become \emph{immutable} by a special operation.

\emph{Slices} of an array can be referenced and be used as operands of some operations.
For example, the Sieve of Eratosthenes program from the previous section can be rewritten using slices:

{\footnotesize
\begin{verbatim}
1.  val flags = [SieveSize : 1];
2.  var i, prime, count = 0, SieveSize = 8191;
3.  for (i = 0; i < SieveSize; i++)
4.    if (flags[i]) {
5.      prime = i + i + 3;
6.      flags[i + prime:SieveSize:prime] = 0;
7.      count++;
8.    }
9.  putln (count);
\end{verbatim}
}

Line 6 in the above example contains an array slice. This slice refers to elements
at indices starting from {\tt i+prime} up to {\tt Sievesize} (exclusive)
with step {\tt prime}.

While array elements are referenced by integer indices,
Dino also includes the associative array type \emph{table}
whose elements are referenced by \emph{keys} that can be arbitrary values. 

Table values can be built using the {\tt tab []} constructor:

{\footnotesize
\begin{verbatim}
1. tab ["string key" : 10.0
2.      [1, 2] : 2,
3.      tab [3, 4] : [1, 2]]
\end{verbatim}
}

The table element with value {\tt 10.0} is stored under a string key,
while an array and another table are used as keys for the table elements in lines 2 and 3 respectively.
As in the case of arrays,
tables can be mutable or immutable, and elements of mutable tables can
be modified, added and removed.

\subsection{Functions, Closures, and Fibers}

  Functions in Dino are first class values.  In other words, they
can be assigned to variables, passed as parameters, returned as
results from other functions and be stored within data structures.  Here
are some example function declarations and function calls:

{\footnotesize
\begin{verbatim}
 1.  fun even;
 2.  fun odd  (i) {i == 0 ? 0 : even (i - 1);}
 3.  fun even (i) {i == 0 ? 1 : odd (i - 1);}
 4   putln (odd (1000000));
 5.
 7.  filter (fun (a) {a > 0;}, v);
 8.  fold (fun (a, b) {a * b;}, v, 1);
 9.
10.  fun incr (base) {fun (incr_val) {base + incr_val;}}
\end{verbatim}
}

According to Dino scope rules, a declaration should be present before any usage of the
declared identifier. Thus, in the case of mutually recursive functions, we need to include
a forward declaration. Lines 1-4 of the above example provide an illustration.

Lines 7 and 8 provide an example of the use of \emph{anonymous} functions.  We pass
the anonymous functions as parameters to the {\tt filter}
and {\tt fold} functions\footnote{Functions {\tt filter} and {\tt fold} are defined in the standard
Dino environment.}.

Function values in Dino always exist within a context, as the
functions can refer to outside declarations. A function value considered along with its
context is called a \emph{closure}. The last line in the above example
illustrates the use of closures.  The function {\tt incr} returns an anonymous
function which in turn returns the sum of its parameter {\tt incr\_val} and the parameter {\tt
base} given to the corresponding call of {\tt incr}.

A \emph{fiber} in Dino is a function which executes concurrently from
its point of invocation.  A fiber call being executed is called a
\emph{thread}.  Here is an example:

{\footnotesize
\begin{verbatim}
1. fiber t (n) {for (var i = 0; i < n; i++) putln (i);}
2. t(100); // the following code will not wait for t finish
3. for (var i = 0; i < 1000; i++) putln ("main", i);
\end{verbatim}
}

To synchronize threads, Dino provides a basic {\tt wait} statement:

{\footnotesize
\begin{verbatim}
    wait (cond) [stmt];
\end{verbatim}
}

Thread is concurrent but not parallel. In other words, the multithreading
is implemented by so-called \emph{green threads}, with the Dino interpreter alternately executing statements from different threads\footnote{The multithreading system in Dino
has existed without any change since its very first version,
as the original application of Dino (scripting dinosaur movements in a simulation)
did not require any more complex form of parallelism.
There are plans
to make threads parallel, e.g. by implementing them as OS-level threads.
This will likely change the semantics of Dino threads in the future.}.
Simple statements, e.g. statements that do not contain function calls, are atomic, i.e. cannot be interrupted by another thread.

\subsection{Classes and Objects}

Some programming languages attempt to unify the functional and
object-oriented approaches.  For example, in Scala a function is just
another kind of object,
 and a function call invokes the method {\tt apply} on the
corresponding object.

Dino takes a bit of a different approach to unifying the two concepts.
  A class in Dino is defined as 
a special
kind of function which returns an entity called a \emph{block instance}\footnote{We prefer to use the
term \emph{block instance} instead of the widely used `activation record' or `stack frame' 
as an activation record is usually allocated on the call stack, whereas a Dino block 
instance is allocated in the heap.} representing the created
object. The inner declarations of the block instance are publicly visible by default.
Here are some example class declarations:

{\footnotesize
\begin{verbatim}
1.  class num (i) {fun print {put (i);}}
2.  class binop (l, r) {
3.    fun print_op;
4.    fun print {l.print(); print_op (); r.print ();}
5.  }
\end{verbatim}
}

Line 1 contains the declaration of a class {\tt num} with one function {\tt print} and one public variable {\tt i} whose value is initialized by the call to the class.  Lines 2 to 5
contain the declaration of an abstract class {\tt binop} with two variables {\tt
l} and {\tt r}, a defined function {\tt print} and a function {\tt print\_op} which declared but not yet defined.

Dino has a powerful class/function composition operator
\emph{use} which can be used to \emph{inlay} declarations from one class into another. 
This operator can be used to emulate \emph{inheritance} (including multiple inheritance), \emph{traits}, and
\emph{dynamic dispatching}. 

Here is a continuation of the above example:

{\footnotesize
\begin{verbatim}
6.  class add (l, r) {
7.    use binop former l, r later print_op;
8.    fun print_op {put (" + ");}
9.  }
\end{verbatim}
}

Line 7 contains a \emph{use}-clause which inlays the
declarations (functions and variables) from class {\tt binop}, making them available within the scope of {\tt add}. 
The \emph{former}-clause overrides the {\tt l} and {\tt r} variables from {\tt binop}
with the {\tt l} and {\tt r} variables defined earlier in the declaration of {\tt add}.
Likewise, the \emph{later}-clause overrides the {\tt print\_op} function inlaid from {\tt binop} with the {\tt print\_op} function defined later in the declaration of {\tt add}.

The \emph{use}-clause provides a safe and powerful way to support object
oriented programming.  It has the following semantics:

\begin{enumerate}
\item Declarations from the class mentioned by the \emph{use}-clause are inlaid into the current class.
\item Declarations from the current class that occur before the \emph{use}-clause override inlaid declarations mentioned in the \emph{former}-clause.
\item Declarations from the current class that occur after the \emph{use}-clause override inlaid declarations mentioned in the \emph{later}-clause.
\item Declarations in the current class must \emph{match} any inlaid declarations that they override.\footnote{An
        exact definition of what it means for two declarations to match
        is not given here. An example
        of matching declarations is a pair of functions with the same number of parameters.}
\item A declaration from the original class mentioned by the \emph{use}-clause can be \emph{renamed} and thus made available in the current class instead of simply being overridden.
\end{enumerate}

A class that references another class via a \emph{use}-clause becomes a \emph{subtype} of the referenced class. The subtyping relation is transitive.
To test subtyping of a class or object, the standard function \emph{isa}
can be used\footnote{Pattern matching can also be used for this purpose.}:

{\footnotesize
\begin{verbatim}
1.  isa (add, binop);
2.  isa (add (num (1), num (2)), binop);
\end{verbatim}
}

Dino provides syntactic sugar for declaring a singleton, which functions as
an abbreviation for declaring an anonymous class, creating an object of the class
and assigning it to a variable declared using \emph{val}:

{\footnotesize
\begin{verbatim}
1.  obj coords {
2.    var x = 450, y = -200;
3.  }
\end{verbatim}
}

\subsection{Pattern Matching}

Pattern matching is a useful instrument for distinguishing and decomposing
values and extracting their parts into variables.
Dino's pattern matching mechanism can
be used in a variable declaration or in a \emph{pmatch-statement}.

The following is an example of pattern matching in a declaration:

{\footnotesize
\begin{verbatim}
1. try {
2.   var [a, b, 2, ...] = v;
3.   ...
4. } catch (patternmatch) {
5.   putln ("your assumption is wrong");
6. }
\end{verbatim}
}

The declaration on line 2 checks that the value of {\tt v} is an array of at
least 3 elements and that the third element of {\tt v} is equal to 2.  If this
is not the case, an exception will occur.
The values of the first and second elements of {\tt v} will be assigned to newly declared variables {\tt a} and {\tt b} respectively.

The following is an example of pattern matching in a pmatch-statement:

{\footnotesize
\begin{verbatim}
1. pmatch (v) {
2.   case [...]: putln ("array"); continue;
3.   case [a, ...]:
4.     if (a == 0) break;
5.     putln ("array with non-zero 1st element");
6.   case node (v) if v != 0:
7.     putln ("object of class node with nozero value");
8.   case _: putln ("anything else");
9. }
\end{verbatim}
}

The pmatch-statement tries to match
a value with the patterns in the case clauses and
executes the code corresponding to the first matched pattern.  The scope
of variables defined in the pattern extends to statements within the case clause.
A \emph{continue} statement within a case clause means that pattern matching should resume and continue to the subsequent case clauses. A \emph{break} statement exits the pmatch
statement.  There is an implicit break at the end of each case clause.

The above example illustrates the possibilities of the pmatch statement, but
the program used for the example is artificial. The following is a more
realistic program using classes and pattern matching:

{\scriptsize
\begin{verbatim}
 1. class tree {}
 2. class leaf (i) {use tree;}
 3. class node (l, r) {use tree;}
 4. fun exists_leaf (test, t) {
 5.   pmatch (t) {
 6.     case leaf (v): test (v);
 7.     case node (l, r):
 8.       exists_leaf (test, l) || exists_leaf (test, r);
 9.   }
10. }
11. fun has_odd_leaf (t) {
12.   exists_leaf (fun (n) {type (n) == int && n % 2 == 1;}, t);
13. }
\end{verbatim}
}

\subsection{Standard Environment}

The Dino standard environment contains a variety of built-in functions
(input/output, higher order functions, regexp functions, functions for
array/table manipulations, etc.) and classes (mostly describing
various exceptions).

This environment is small in comparison with the standard libraries of popular dynamic
languages. The creation of a full set of standard libraries requires a lot of
efforts and has not been a priority as Dino was more of a research project for a long time.  That said, there
is one unique feature in the Dino standard environment: a
predefined class for the creation of parsers for programming languages or
natural languages.

The class implements an enhanced \emph{Earley parser} algorithm with
\emph{simple syntax directed translation}.
It can parse according to ambiguous grammars, producing either a compact representation of \emph{all} possible parse trees or a \emph{minimal cost} parsing tree when costs are assigned to the grammar rules. 
The parser can perform syntax recovery, finding the \emph{minimal}
number of ignored tokens which produces a correct abstract syntax tree (AST).  It is also fairly fast, being capable of parsing about 250,000 lines of C
language code per second on modern CPUs\footnote{The implementation of Earley's parser used in Dino achieves 255,700 lines per second when parsing a preprocessed 67,000 line C file on 4.2GHz Intel i7-4790K. Peak memory usage for parser internal data was about 3.9MB.}.

Here is an example where the Earley's parser built into Dino is used to parse a tiny programming language:

{\scriptsize
\begin{verbatim}
expose yaep.*;
val grammar = "TERM ident=301, num=302, if=303,
                    then=304, for=305, do=307, var=308;
  program = program stmt                     # list (0 1)
  stmt = ident '=' expr ';'                  # asgn (0 2)
       | if expr then stmt else stmt         # if (1 3 5)
       | for ident '=' expr expr do stmt     # for (1 3 4 6)
       | '{' program '}'                     # block (1)
       | var ident ';'                       # var (1)
       | error
  expr = expr '+' factor                     # plus (0 2)
  factor = factor '*' term                   # mult (0 2)
  term = ident                               # 0
       | '(' expr ')'                        # 1";
val p = parser ();       // create an Earley parser
p.set_grammar (grammar); // set grammar
fun syntax_error;        // forward decl of error reporting func
val abstract_tree = p.parse (token_vector, syntax_error);
\end{verbatim}
}

\begin{center}
\begin{figure*}[ht]
\begin{center}
\includegraphics[scale=0.5]{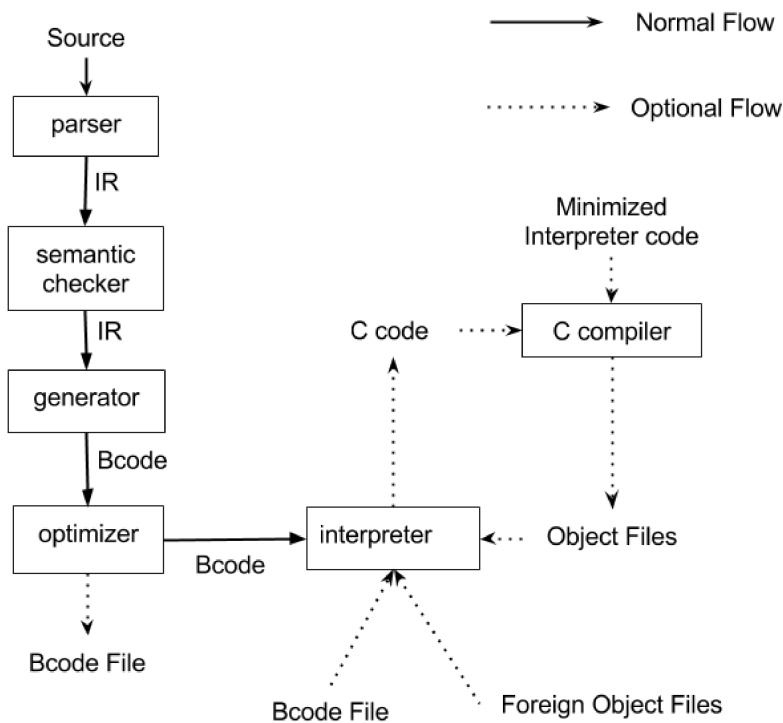}
\end{center}
\caption{Overall Dino data flow}
\label{fig:Dino-flow}
\end{figure*}
\end{center}

\section{Implementation}
  Unfortunately, the article size does not permit a description
of all design decisions, algorithms and optimizations in full detail.
Only a high level overview is given, and only some important elements
of the implementation are described in more detail.
  
  Figure \ref{fig:Dino-flow} represents the general structure of the current Dino
implementation\cite{Makarov} and the flow of data through it.

  Dino interpreter can be used in one of two modes.  The first mode is a read-eval-print 
loop (REPL) where a completed statement undergoes all
processing stages including execution. 
The second mode is the standard batch mode, where the interpreter
first processes all program files to generate byte code of the entire
program and then executes the byte code.  Program byte code can
be saved in a readable form, modified and read back for execution.

  The interpreter includes a function-level JIT compiler which is implemented
with the aid of a C compiler.

  The Dino program can load and execute PIC object files created from
C code using a foreign function interface.

  The interpreter uses a memory heap with garbage collection.  The
heap can be automatically extended by demand.  The interpreter performs a
simple escape analysis to transform heap allocations into stack ones.
The garbage collector uses a combination
of the mark-and-sweep and the fast mark-and-copy strategies
to prevent heap fragmentation and
decrease program memory requirements \cite{GCH}.

  The first implementation of associative tables in Dino was based on classical hash
tables with buckets.  The implementation was later changed to
resizable hash tables without buckets\footnote{The hash function used in Dino is based on MurMur hashing \cite{Appleby}.}. 
Conflict resolution is
achieved by secondary hashing.  Such an approach permits to decrease
pointer chasing and results in more compact hash table
representations and better data locality.
With the right choice
of maximal load factor
for the hash table, the implementation using secondary hashing compared to the implementation using
buckets can be 1.5 times faster on modern processors\footnote{Both hash table implementations provide the same
  functionality and approximately the same overall table sizes and
  rate of growth.
  Hashing benchmarks from the old version of the computer language
  shootout were used for the comparison \cite{Benchmarks}.}.
The hash table implementation based on secondary hashing was later adapted for use in the GCC compiler.

   The Dino interpreter employs numerous optimizations to improve performance.
Below we discuss optimizations which are important from the performance point of view
or widely described in research literature.

\subsection{Byte Code Dispatch}

  Although a lot of articles and discussions about dynamic language
implementations are focused on \emph{byte code dispatch}, in our experience 
the choice of dispatch
method was found to be relatively unimportant for the performance of the Dino implementation.

  In the \emph{direct threaded code} method, each byte code
instruction contains the address of the code responsible for execution of
the next byte code instruction \cite{Bell} \cite{Ertl}.  For example, the Ruby VM
uses direct threaded code by default \cite{Sasada}.
Although the code required for dispatch is very simple (read the address and jump to it), implementing the direct threaded strategy requires a nonstandard extension to the C language which allows labels to be treated as values \cite{GCC}. The use of this extension prevents the compiler from performing some optimizations on the interpreter code (e.g. predictive commoning).

  The \emph{direct dispatch} method is based on a standard C switch
statement. Each case of the switch statement contains an implementation of a different byte code instruction. Dino implementation
experience shows that a \emph{proper} direct dispatch implementation is
more efficient on modern architectures. By `proper', we mean that the compiler
should be forced to generate optimized code for the switch statement
if the number of all byte code intructions is less than 256\footnote{For example, it is possible to force the GCC compiler to omit range checking and zero extension of the opcode value. The resulting code for x86/x86-64 has less than half the instructions compared to the code generated for a switch statement by default.}. The optimization results in the code for direct dispatch being compiled to a number of machine instructions comparable to that of the machine code for direct threaded dispatch.

  To take advantage of hardware branch prediction, some authors proposed
to replicate instructions or to make a direct switch
at the end of each code responsible for executing a byte code instruction \cite{Casey}.  This
might work well when we have a small benchmark that takes advantage of the hot path,
but in most real programs the bytecode will generally have different instructions adjacent to one another. 
The larger and more complex the benchmarks, and the more different types of
byte code instructions the VM provides, the worse this approach will perform.
The performance rapidly becomes worse than that of direct dispatch.

  In early Dino implementation, the use of a proper direct dispatch implementation improved
code performance for a simple loop benchmark up to 3\% (on a modern
x86-64 architecture) compared to the next-best direct threaded
dispatch method.  Unfortunately, the latest Dino implementation uses more
than 256 byte code instructions.  Therefore we started to use the direct threaded code
method as the best possible dispatch method for such number of byte
code instructions.

  A much greater performance improvement can be reached by focusing on other
issues than the choice of byte code dispatch.  The most important aspect of the design
is choosing the right byte code architecture for efficient execution.

\subsection{Virtual Machine and Bytecode Design}

  \emph{Stack-based} (SB) virtual machines (VMs) are a fairly popular design for
implementation of dynamic language interpreters. For example, the Ruby
VM uses a stack-based architecture \cite{Sasada} \cite{Shaughnessy}.  Stack based byte
code is compact and simple.

The previous implementation of Dino VM used a stack-based design as well.
The current implementation is based on a \emph{register transfer
  language} (RTL).  Changing the virtual machine architecture from a stack-based to
a register-transfer-language VM architecture improved code
performance up to \textbf{3} times,
which is bigger than the speedup reported
in \cite{Shi}\footnote{There are many RTL design details
  which could influence performance    
  and may account for the difference in speedup.
  For example, the addressing system in Dino bytecode
  allows an instruction to address both local and temporary variables as well as
  variables in global scope.}.
  
  The main advantage of RTL is a reduced number of instructions, as one RTL instruction can correspond to 3 or more SB
byte code instructions. In many cases, the reduced number of instructions decreases 
the overhead of byte code dispatch and reduces the amount of operand shuffling and
type checking. Moreover, there is greater scope for 
applying compiler optimizations to the bytecode interpreter itself
if it is based on an RTL rather than an SB design.

  The benefit of RTL is even greater when we take into account the ability to
perform optimizations on the byte code, including instruction combining, instruction specialization, and colour based variable allocation. These optimizations are
very difficult or even impossible to implement for SB virtual machines.

  The current Dino byte code format includes multi-operand instructions
with 1-5 operands (usually 3), control flow instructions (blocks, branches, calls, etc.),
as well as instructions representing declarations such as vdecl (variable declaration)
and fdecl (function, class, and fiber declarations).

  Dino byte code has two representations: an in-memory format for execution
and a human-readable representation which can be output, modified, and input
back by the interpreter for further execution.

  The Dino code

{\footnotesize
\begin{verbatim}
  var i, n = 1000;
  for (i = 0; i < n; i++);
\end{verbatim}
}

  is compiled to byte code which includes the following segments (given in human-readable representation):
 
{\scriptsize
\begin{verbatim}
0 block fn="ex.d" ln=1 pos=1 next=730 vars_num=29 tvars_num=3
...
372 vdecl ... ident=i ident_num=268 decl_scope=0 var_num=27
373 vdecl ... ident=n ident_num=269 decl_scope=0 var_num=28
...
788 ldi ... op1=28 op2=1000 // 28 <- i1000
789 ldi ... next=791 op1=27 op2=0 // 27 <- i0
790 btltinc ... next=792 op1=27 bcmp_op2=28 bcmp_res=29 pc=790
791 btlt ... op1=27 bcmp_op2=28 bcmp_res=29 pc=790
792 bend ...
\end{verbatim}
}

\subsection{Optimizations}

  Choosing the right byte code representation is
crucial, but there are other techniques 
which are important for attaining good performance.

\subsubsection{Byte Code Combining}

  Reducing the number of executed instructions provides a performance advantage,
due to reduced dynamic dispatch overhead and an
increased scope for applying compiler optimizations when building the Dino bytecode interpreter. The use of RTL representation is one way of achieving this.

  The same result can also be achieved by creating new bytecode instructions and using them to 
replace frequently executed chains of operations. Such an optimization
is performed by the Dino interpreter. An analogous approach can be found in
\cite{Abdelrahman}, where this optimization is called \emph{concatenation}, or in
\cite{Casey}, where it is called \emph{combining instructions into superinstructions}.

The following example illustrates how the Dino interpreter transforms byte code
corresponding to an empty for loop:

{\scriptsize
\begin{verbatim}
  label: addi op1, op1, i1; lt res, op1, op2; bt res, label =>
  label: addi op1, op1, i1; blt res, op1, op2, label =>
  label: btltinc op1, op2, i1, res, label
\end{verbatim}
}

  Byte code combining improves the performance of the model benchmarks described in section \ref{model_benchmarks} by about 1.5 times.

\subsubsection{Pure Function Calls}

  A function is called pure if it has no side effects and
its result always depends only on its argument values.
  When a pure function is called many times with
the same arguments, we don't need to calculate its result again.

  Dino interpreter can save values of arguments and corresponding
results of pure function calls and then reuse the results instead of
performing the calculations again.  This optimization can improve the performance
of some benchmarks such as factorial or Fibonacci by an order of magnitude.

\subsubsection{Just-in-time Compilation}

  Performance of frequently executed code can be improved by just-in-time
compilation (JIT) to machine code. There are a number of approaches to selecting which unit of 
code to compile. The most frequently used approach is \emph{trace-based} selection.
Examples of trace-based JIT implementations include the PyPy implementation of Python and 
most widely used JavaScript implementations.
JIT compilation can be implemented through a specialized JIT framework
(e.g. JVM\footnote{If JVM is used for the implementation, the bytecode obviously has to be Java bytecode.})
or by using more general compiler frameworks such as LLVM or the GCC JIT plugin \cite{Lattner} \cite{Malcolm}.

  The Dino interpreter uses function-level JIT.  Function compilation is
triggered by the first call.  The function's byte code is translated
into C function code.
For each type of byte code instruction there is a corresponding
C inline function. When building the Dino interpreter, these functions
are copied to a C-preprocessed file, which is included in the generated C
code during JIT compilation\footnote{The C inline functions implementing the byte code instructions
are declared \emph{static}. Thus, GCC does not
  optimize or generate separate machine code for these functions;
  they are inlined only if they are used.
  This considerably increases the speed of JIT compilation.
  Another important factor that increases the compilation speed
  is the fact that the preprocessed file contains only declarations that
  are used by the inline functions, rather than the full set of declarations
  from the Dino interpreter.
}. Thus, the
C translation of a function's byte code consists mostly of calls to
these functions.
A C compiler is called with the generated C code provided through a UNIX pipe.  The
C compiler produces a position independent code (PIC) object file which is
stored in the operating system's standard temporary directory\footnote{These days, UNIX systems usually store their temporary directory in memory.}.
After successful compilation, the interpreter loads the
object file and calls the compiled function.

  This approach has many advantages.  First of all, it results in a
simple and portable JIT implementation.  In the Dino interpreter, the JIT implementation
required less than 100 new lines of C code, and can be used with any native C compiler.
As the C language has a stable and well-defined standard,
Dino's portable JIT implementation will continue to work in the future.
On the other hand, if we had used the special-purpose interfaces provided
by LLVM and GCC for JIT compilation, we would potentially need to update the interpreter 
in order to be compatible with future compiler versions.

Machine code generation by passing a C source file to the GCC frontend gives only a slight performance penalty compared to generation
using the GCC JIT plugin\footnote{Experiments with the JIT plugin benchmark from
  GCC's testsuite show that compilation using the GCC JIT plugin is about
  20\% faster than compiling from source code on a Xeon E5-2697 v3
  machine.}.  In fact, JIT compilation of a small Dino function takes only about
50-70ms using GCC with -O3 on modern Intel CPUs\footnote{A heapsort function
 was used for the measurement on a Xeon E5-2697 v3 machine.}.

\subsubsection{Type-inference and Byte code Type Specialization}

  Dino is a dynamically typed programming language.  Nevertheless, the types of
many expressions and operands can be recognized during compilation time.

  Unfortunately, even the peak optimization modes of industrial C compilers used for JIT are not powerful enough to figure out the types of
byte code operands and to remove unnecessary operand type checking.

  Type recognition (\emph{inference}) with byte code specialization
can help significantly to improve JIT performance.  For example,
when we know the types of operands,
byte code specialization can transform the general {\tt add} instruction
into an integer variant {\tt iadd} or a floating point
variant {\tt fadd}.  On our model benchmark,
qualitative
byte code specialization improves JIT performance about 4.5 times.
With the addition of combined and specialized instructions,
the number of different byte code instructions
in Dino is 254.

  The type inference algorithm in Dino consists of the following major steps:

\begin{enumerate}
     \item Building an \emph{all-program} control flow graph (CFG) consisting of basic blocks and control flow edges connecting them.
    \item Calculating available results of byte code instructions -- this is a forward data-flow problem on the CFG.
    \item Using the availability information, building def-use chains
connecting possible operands and results of byte code instructions and variables.
    \item Calculating the types of byte code instruction operands and results -- this is a forward data-flow problem on the def-use graph.
\end{enumerate}

  This algorithm is capable of recognizing the types of different usages of a variable
even if the variable holds values of different types during an execution.  This is a major difference from type inference in statically typed languages.

  Dino has a number of language features and edge cases (such as higher order functions, closures, threads and the possibility of accessing an uninitialized variable) that make comprehensive type inference difficult.
Therefore we do not
recognize all operand types which it is theoretically possible to recognize.

\subsubsection{Code Reuse in Class Composition}

  A naive implementation of the class composition operator {\tt use} that duplicates declarations
can create a lot of additional code.  Therefore special attention during the
implementation of Dino and the design of its byte code was paid to avoiding this
problem.  In fact, no code duplication is produced by the current
Dino implementation of operator {\tt use}.  Byte code corresponding to a function defined in one class
and inlaid into another class is always reused.

\subsubsection{Optimization Hints}

After analyzing his program's behaviour, the programmer usually has the best knowledge
about what should and should not be optimized.
Therefore Dino allows the programmer to attach optimization hints to functions. Currently,
Dino defines hints to request \emph{inlining}, \emph{JIT compilation}, or \emph{pure
function optimization}.

Let us look at how optimization hints can be used to tune the performance of the
`meteor-contest'
benchmark from the new version of the computer language shootout \cite{Meteor}.
Using the built-in Dino profiler
(option {\tt -p}), we obtain the following output:

{\scriptsize
\begin{verbatim}
** Calls *** Time **** Name *********************************
  761087        0.43  --  search1: "meteor.d": 229
  561264        0.07  --  ctz: "meteor.d": 28
    1260        0.01  --  GoodPiece: "meteor.d": 37
     ...
                0.51  --  All Program
\end{verbatim}
}

We know that {\tt ctz} is a small function and by inlining it we can
remove significant call overhead.  Function {\tt search1} is larger, and it
may be better to compile it into native code using the JIT compiler.  After
adding hints {\tt !inline} and {\tt !jit} to the corresponding functions,
we obtain the
following profiler output:

{\scriptsize
\begin{verbatim}
** Calls *** Time **** Name *********************************
  761087        0.15  --  search1: "meteor.d": 229
     ...
       0        0.00  --  ctz: "meteor.d": 28
     ...
                0.17  --  All Program
\end{verbatim}
}

\subsection{Used Tools and Libraries}

The development of dynamic languages is usually an ongoing process.
Thus, it is important to develop tools that make it easier to implement
and modify the language interpreter. The COCOM toolset provides several
tools that are used by the Dino implementation.
%

The most useful one is Sprut, a compiler of internal representation (IR)
descriptions.  The description is given in an object oriented format.  The
Sprut compiler generates code to create, access, modify,
traverse, read and write the described IR format.  Additionally, in debug mode Sprut
can generate code to check constraints and relations given in the IR description.

There are four IR descriptions included in the Dino implementation.
The first is a high-level IR used for semantic checking.
The second is the Dino bytecode format. The third format
is a part of the foreign function interface and is used to specify
run-time data which should be visible to external C functions.
The last format describes all run-time data used by the Dino interpreter.

The following are some excerpts from the semantic IR description:

{\footnotesize
\begin{verbatim}
...
%abstract ir_node :: %root;
%abstract generic_pos :: ir_node
%other
  pos : position_t
;
%abstract expr :: generic_pos;
%abstract operation :: expr;
%abstract binary_operation :: operation
%other
  left_operand, right_operand : expr
;
%abstract unary_operation :: operation
%other
  operand : expr
;
mult :: binary_operation;
typeof :: unary_operation;
...
\end{verbatim}
}

The following are excerpts from the bytecode description:

{\footnotesize
\begin{verbatim}
...
%abstract icode :: %root;
%abstract bcode :: icode
%other
  next : bcode
  info : %root
;
%abstract op1 :: bcode
%other
  op1 : int_t
;
%abstract op2 :: op1
%other
  op2 : int_t
;
%abstract op3 :: op2
%other
  op3 : int_t
;
%abstract op3i :: op3;
add :: op3;
addi :: op3i;
\end{verbatim}
}

Coding a parser manually can complicate language development.
Therefore the Dino parser is implemented using a parser generator
called Msta.  Msta is a superset of YACC/Bison, supporting
LALR(k)/LR(k) grammars with any fixed $k \geq 0$.  Msta also includes more sophisticated error
recovery algorithms and implements many useful
optimizations\footnote{Msta can extract regular or LALR portions of an LR
  grammar and generate the corresponding portion of the parser
  as a DFA or LALR stack machine.}.
Msta generates about 25\% faster parsers than Bison\footnote{When a
preprocessed 67,000 lines C file is parsed on 4.2GHz Intel
i7-4790K, Msta is about 26\% and 36\% faster than Bison and Byacc correspondingly.}.

The third COCOM tool used by Dino is Shilka, a generator of fast keyword recognizers.
There are several components of the Dino implementation that need to perform recognition
of reserved identifiers, e.g. Dino keywords, bytecode instruction
names, and bytecode instruction fields.  Shilka's functionality is analogous to that of the GNU
gperf package but its implementation uses minimal pruned O-tries.  This approach makes Shilka
up to 40\% faster than gperf\footnote{Shilka on i7-4790K was 38\%
  faster than gperf when recognizing 38 million C language keywords
  out of 120 million identifiers.  The sequence of identifiers was taken from a
  large C source file.}.

The COCOM toolset also includes a library called Ammunition. This library contains
different packages useful for compiler/interpreter implementations, such as
source position handling, error reporting, memory allocation handling,
hash tables, sparse sets, Earley parser etc.  The Dino implementation
actively uses most of these.

Additionally, the Dino implementation uses the GMP (multi-precision
integer arithmetic), iconv (different encodings support), and ONIGURUMA regex libraries.

\section{Effect of implementation decisions and optimizations}
\label{model_benchmarks}

  One of the reasons for wide adoption of dynamic languages is the
high-level nature of their operation.  Usually, performance of low-level
operations is the weak point of dynamic language implementations.

  Thus, we have chosen a benchmark which allows us to study the
performance of low-level operations. The nested loops benchmark from the old
version of the
computer language shootout could serve well for
this purpose \cite{Benchmarks}.  The Dino version of this benchmark consists of the following code:

{\scriptsize
\begin{verbatim}
fun main {
  var n = argv [0] < 1 ? 1 : int (argv[0]);
  var a, b, c, d, e, f, x = 0;

  for (a = 0; a < n; a++)
    for (b = 0; b < n; b++)
      for (c = 0; c < n; c++)
        for (d = 0; d < n; d++)
          for (e = 0; e < n; e++)
            for (f = 0; f < n; f++)
              x++;
  putln (x);
}
main ();
\end{verbatim}
}

  We ran this benchmark using different Dino VMs and optimizations on
three platforms: x86-64, AARCH64, and Powerpc64.
The machines used for testing
were a 4.2GHz Intel 4790K, a 2.4Ghz Applied Micro AARCH64 X-gene, and
a 3.2GHz Power8 running Linux Fedora Core 21.  Version 4.9.2 of GCC was used for
JIT compilation.  Each test was run 3 times on an unloaded
machine and the minimal
time was taken. The same command line argument (25) was passed to the Dino script in all runs.
Table \ref{tab:optimizations} contains the CPU times of different runs of the benchmark in seconds.

\begin{table}[h]
\begin{center}
\begin{tabular}{|l|r|r|r|}
\hline

                          & Intel         & Applied Micro    & IBM    \\
                          & Haswell       & X-gene           & Power8 \\ \hline
Stack based VM            &       6.51    & 34.23            & 24.99  \\ \hline
RTL VM                    &      2.12     & 13.08            & 8.84   \\ \hline
The above + code combining    &      1.38     & 6.92             & 4.69   \\ \hline
The above + JIT               &      0.61     & 2.13             & 2.20   \\ \hline
The above + type inference   &      0.14  & 0.24             & 0.11 \\
and byte code specialization &                &                  &    \\ \hline
\hline

\end{tabular}
\end{center}
\caption{Performance effect of design decisions and optimizations (CPU
time given in seconds).}
\label{tab:optimizations}
\end{table}

\section{Comparison with a few dynamic language implementations}

  To demonstrate overall Dino performance in comparison with popular
implementations of Ruby, Python and Javascript on a wider set of benchmarks,
a subset of the old version of the computer language shootout was used \cite{Benchmarks}.

\begin{center}
\begin{table*}[ht]
\begin{center}
\begin{tabular}{|l|r|r|r||r|r|r|r|r|r|r|r|r|r|r|}
\hline
  
        &Loop   &Hash&Fact   &Fib    &Except&Method&Object&Sieve&Sort &Stat.&Random &Thread&Start&Compile \\ \hline
Dino    &1.0    &1.0 &1.0    &1.0    &1.0   &1.0   &1.0   & 1.0 &1.0    &1.0  &1.0    &1.0   & 1.0 & 1.0   \\
Python  &416    &2.2 &156    &483    &9.0   &7.8   &6.0   &54.6 &9.4    &3.6  &34.6   &190   &26.3 &  3.9  \\
Ruby    &237    &2.8 &44.6   &148    &4.6   &1.9   &2.5   &  8.3&3.1    &4.6  &13.4   &194   &25.5 &  2.0  \\
PyPy    &23.7   &0.5 & 0.6   &  68   &0.5   &0.3   &0.1   &  6.9&1.2    &1.5  &  1.3  &76.5  &22.1 &21.3   \\
JS      &204    &1.1 &52.4   &167    &   -  &   -  &  -   &  7.8&0.5    &  -  &  0.7  &  -   &  0.8&  1.9  \\
Scala   &10.7   &1.1 & 3.3   &114    &8.9   &1.4   &0.8   &  2.8&0.5   &7.1  &  1.3  &  -   &377  &  -     \\ \hline

\end{tabular}
\end{center}
\caption{Performance of Dino vs. performance of other language implementations on Intel Haswell.  All execution times of benchmark programs are given relative to the execution time of the corresponding Dino version.}
\label{tab:x86-64}
\end{table*}
\end{center}

\begin{table*}[t]
\begin{center}
\begin{tabular}{|l|r|r|r||r|r|r|r|r|r|r|r|r|r|r|}
\hline

        &Loop   &Hash   &Fact   &Fib    &Except&Method&Object&Sieve&Sort   &Stat.&Random &Thread&Start&Compile\\ \hline
Dino    &1.0    &1.0    &1.0    &1.0    & 1.0  & 1.0  &1.0   & 1.0 &1.0    &1.0  &1.0    & 1.0  & 1.0 &1.0    \\
Python  &222    &0.5    & 68.8  &1116   &12.4  & 4.7  &2.8   &40.3 & 5.3   &2.5  &15.5   &149   &246  &2.6    \\
Ruby    &170    &2.8    &32.3   &  542  &  6.0 & 1.7  &1.4   & 8.5 & 3.7   &3.8  &13.1   &118   &655  &1.6    \\
JS      &282    &1.4    &31.6   &  471  &   -  &  -   &  -   & 16.2& 2.5   &  -  & 6.5   & -    & 1.0 &0.5    \\
\hline

\end{tabular}
\end{center}
\caption{Performance of Dino vs. performance of other language
  implementations on Applied Micro X-gene.  All execution times of benchmark programs are given relative to the execution time of the corresponding Dino version.}
\label{tab:X-gene}
\end{table*}

\begin{table*}[t]
\begin{center}
\begin{tabular}{|l|r|r|r||r|r|r|r|r|r|r|r|r|r|r|}
\hline
  
        &Loop   &Hash   &Fact   &Fib    &Except&Method&Object&Sieve&Sort   &Stat.&Random &Thread&Start&Compile\\ \hline
Dino    &1.0    &1.0    &1.0    &1.0    &1.0   &1.0   &1.0   &1.0  &1.0    &1.0  &1.0    &1.0   &1.0  &1.0    \\
Python  &25.3   &0.5    &65.9   &592    &12.9  &4.3   &6.1   &25.1 &4.3    &2.0  &6.9    &166   &62.5 &2.4    \\
Ruby    &17.6   &1.7    &43.0   &390    &20.0  &1.6   &3.1   &7.6  &4.2    &5.2  &11.4   &62.8  &46.1 &1.5    \\
JS      &27.0   &3.0    &53.8   &453    &   -  &   -  &  -   &14.2 &2.8    &  -  &4.6    &  -   &2.7  &0.7    \\
\hline

\end{tabular}
\end{center}
\caption{Performance of Dino vs. performance of other language
  implementations on Power7. All execution times of benchmark programs are given relative to the execution time of the corresponding Dino version.}
\label{tab:power7}
\end{table*}


This subset includes the following benchmarks:
\begin{itemize}
\item loop -- execution of empty loop body.
\item hash -- associative table operations.
\item fact -- recursive factorial function.
\item fib -- recursive Fibonacci function.
\item except -- exception handling.
\item method -- object method calls.
\item object -- object instantiation.
\item sieve -- Sieve of Eratosthenes algorithm.
\item sort -- sorting using the heapsort algorithm.
\item stat -- calculation of statistical moments.
\item random -- random number generation.
\item thread -- passing data from a producer thread to a consumer thread.
\item start -- an empty program that does nothing.
\item compile -- a program that consists of a very long series of assignment statements.
\end{itemize}

  Test machines used to run the benchmarks were a 3.4GHz i5-4670 (x86-64 Haswell),
a 2.4GHz Applied Micro X-gene (AARCH64), and a 3.5GHz Power7 (Powerpc64)
running Linux Fedora Core 21.

  Language implementations used for the comparison include
the Python-3.3.x interpreter,
the Ruby-2.0.x interpreter, and the JavaScript-1.8.x SpiderMonkey and TraceMonkey
JITs.  Results for the PyPy-2.2.x trace JIT for Python and the Scala-2.10.x
JVM JIT are given only for x86-64 as these systems are not yet implemented
on the AARCH64 architecture\footnote{This can be considered an illustration of the difficulties of porting dedicated JIT systems
to a new architecture.}.  The absence of results for PyPy on
Powerpc64 is due to the same reason.  Finally, although Scala and the JVM are
implemented on Powerpc64, unavailability of Scala on the
machine used for testing prevented us from obtaining the corresponding results.

  Tables \ref{tab:x86-64}, \ref{tab:X-gene}, and \ref{tab:power7}
contain benchmark execution times on x86-64,
AARCH64, and Powerpc64 platforms respectively.  All times are scaled relative to the execution time of the Dino version of the benchmark program.

  The best results for the Dino version of the factorial and Fibonacci benchmarks were
achieved by using pure-function optimization.  The best result
for the Dino version of the random number generator benchmark was achieved by using function inlining.

   Results for some JavaScript benchmark programs are absent because the language
does not support features needed for their implementation, or
because JavaScript versions of these benchmark programs could not be found.
On AARCH64 the result for the JavaScript `hash' benchmark is given with JIT disabled, as JIT-enabled execution failed.

Scala results for the `compile' benchmark are absent, as Scala failed to compile it.

\section{Conclusion and Perspectives}

  There are many venues for future research on Dino. The effect of new optimizations (such 
as improved type inference or fast variable allocation using a linear scan algorithm 
\cite{Poletto}) can be investigated.
We also have plans to use
Dino to research various JIT approaches. One way of improving JIT can be
achieved by maintaining a C compiler process (or a pool of compiler processes) on standby
in order to avoid the overhead of starting up a compiler during JIT compilation of each function.

  A very important research goal is to find improved ways of implementing JIT based
on GCC, as it is the most portable 
compiler infrastructure currently available\footnote{GCC-5.0 includes support for 48 targets in its public repository, while LLVM-3.6 supports only 12 targets.}.  It is
fairly reasonable to doubt that any JIT compiler written from scratch
can achieve the same level of portability.

  A possible way to improve a GCC-based JIT compiler is to investigate which
GCC optimizations actually improve the performance of the C code generated by the JIT compiler. As described earlier, type inference is essential, as GCC is not able to
propagate constants which represent operand types in a dynamic language program,
nor can it remove code for checking operand types\footnote{Conditional constant 
propagation on C code generated by a JIT compiler can be considered to be analogous to
type inference on dynamic language code \cite{Morgan} \cite{Muchnick}.}. Some 
optimizations in GCC, e.g. loop invariant code motion, may also not work as well as 
expected when applied to C code generated by a JIT compiler. This problem could be solved 
by improving the optimizations in GCC or by implementing some additional
byte code level optimizations in the interpreter\footnote{Even if we would need to 
implement the majority of optimizations on the byte code level, GCC will still be useful 
as a non-optimizing but highly portable C compiler.}.

  Another way to improve a GCC-based JIT compiler could be to find the
right combination of GCC optimizations which is fast enough but
still considerably improves the performance of the generated C code.
Such a combination should aim to balance the time of JIT compilation itself
against the execution time of the generated code.
Some machine learning approaches could be
used in order to find the best combination of optimizations
depending on features of the generated code \cite{CompilerDesign}.

  Still GCC-based JIT can be quite expensive for some targets which
have memory constraints or do not support files in memory.  One such
target is CYGWIN environment\cite{Cygwin} where using GCC-based JIT
results in a performance degradation in most cases.  A simple light-weight JIT
implementation specialized for Dino could be a solution.  Comparison
of GCC-based JIT and the light-weight JIT with different points of
view (generated code performance, efforts to implement, used
resources, startup time etc) would be quite an interesting topic for research.
  
There are also plans to develop Dino from a research language into a language which could
be widely used in practice. Features planned for this purpose include
type annotation,  light-weight specialized JIT, native OS thread support for parallelism,
rewritten standard libraries, and a more convenient foreign function 
interface. In addition to native support for parallelism,
a more sophisticated model could be provided for synchronization.

  During the long history of the Dino programming language, a lot of research
has been done.  Part of this research was described in this article,
including the effect of major implementation decisions on performance,
some tools which can be used to simplify the implementation, and ways to
implement a simple, portable, and stable JIT.

  We hope that the results reported in this article could
help to find ways to improve popular dynamic language implementations
to make them faster and more portable, using fewer developer resources.



\bibliographystyle{abbrvnat}

\begin{thebibliography}{}
\softraggedright

\bibitem[Abdelrahman]{Abdelrahman}
B. Vitale and T.S. Abdelrahman. ``Catenation and specialization for
Tcl virtual machine performance'', \emph{Proc. of the Int'l Workshop on
Interpreters, Virtual Machines and Emulators}, pp. 42-50, Washington,
DC, June 2004.

\bibitem[Appleby]{Appleby}
A. Appleby. MurmurHash 2.0,
http://murmurhash.googlepages.com/

\bibitem[Bell]{Bell}
  J. Bell. ``Threaded code'', \emph{Comm. ACM}, vol. 16, no. 6,
  pp. 370-372, 1973.

\bibitem[Benchmarks]{Benchmarks}
  The Great Win32 Computer Language Shootout,
  http://dada.perl.it/shootout/
  
\bibitem[Casey]{Casey}
  K. Casey, M. A. Ertl, and D. Gregg. ``Optimizing indirect
  branch prediction accuracy in virtual machine interpreters.'',
  \emph{ACM Transactions on Programming Languages and Systems},
  vol. 29, no. 6, 2007.

\bibitem[CompilerDesign]{CompilerDesign}
  K. Vaswani. ``Statistcial and Machine Learning Techniques in Compiler Design'',
  ``The Compiler Design Handbook: Optimizations and Machine Code Generation'',
   Second Edition, CRC Press, 2007.
  
\bibitem[Cygwin]{Cygwin}
https://www.cygwin.com/

\bibitem[Ertl]{Ertl}
M.A. Ertl, D. Gregg. ``The behavior of efficient virtual machine
interpreters on modern architectures'', \emph{Euro-Par 2001 Parallel
  Processing},  pp. 403-413, 2001, Springer.

\bibitem[GCH]{GCH}
R. Jones. ``Garbage Collection Handbook: Art of Automatic Memory Management'',
2011, Chapman and Hall/CRC.

\bibitem[GCC]{GCC} The GNU compiler collection, \emph{Labels as Values},
https://gcc.gnu.org/onlinedocs/gcc-5.1.0/gcc/Labels-as-Values.html\#Labels-as-Values
  
\bibitem[Lattner]{Lattner}
Chris Lattner and Vikram Adve.  \emph{LLVM: A compilation
  Framework for Lifelong Program Analysis and Transformation},
International Symposium on Code Generation and Optimization, March
2004.

\bibitem[Makarov]{Makarov} Vladimir Makarov, The Programming Language Dino,
https://github.com/dino-lang/dino
  
\bibitem[Malcolm]{Malcolm} David Malcolm, GCC JIT wiki page,
https://gcc.gnu.org/wiki/JIT
  
\bibitem[Meteor]{Meteor}
  Description of meteor contest benchmark,
  http://benchmarksgame.alioth.debian.org/u32/performance.php?test=meteor
  
\bibitem[Morgan]{Morgan} Robert Morgan.  ``Building an Optimizing
  Compiler'',  Digital Press, 1998.

\bibitem[Muchnick]{Muchnick} S. Muchnick. ``Advanced Compiler Design and
  Implementation'',  Morgan Kaufmann, 1997.

\bibitem[Poletto]{Poletto}
M. Poletto and V. Sarkar, ``Linear Scan Register Allocation'',
\emph{ACM Transactions on Programming Languages and Systems}, 1999,
vol. 21, no. 5, pp. 895-913.
      
\bibitem[Sasada]{Sasada}
K. Sasada. ``YARV: yet another RubyVM: innovating the ruby
interpreter'', in \emph{Companion to the 20th annual ACM SIGPLAN
conference on Object-oriented programming, systems, languages, and
applications}, ACM, New York, NY, USA, pp. 158-159.

\bibitem[Shaughnessy]{Shaughnessy}
Pat Shaughnessy. ``Ruby Under a Microscope: An Illustrated Guide to
Ruby Internals'', No Starch Press, 2013.
  
\bibitem[Shi]{Shi}
Y. Shi, K. Casey, and M. Anton Ertl, and D. Gregg. ``Virtual machine
showdown: Stack versus registers''. \emph{ACM Transactions on Architecture
 and Code Optimization (TACO)}, Volume 4 Issue 4, January 2008
Article No. 2.

\end{thebibliography}


\end{document}